\documentclass{pnastwob} 

\usepackage{graphicx}
\usepackage{amssymb,amsfonts,amsmath}

\begin{document}

\title{Mechanical Surface Waves Accompany Action Potential Propagation}

\author{Ahmed El Hady$^a$\affil{1}{Princeton Neuroscience Institute}\affil{2}{Howard Hughes Medical Institute} 
\and
Benjamin B. Machta$^b$\affil{3}{Lewis-Sigler Institute, Princeton University, Princeton, NJ 08544}}
\contributor{$^a$ email: ahady@princeton.edu, $^b$ email: bmachta@princeton.edu}
\copyrightyear{}
\issuedate{}
\volume{}
\issuenumber{}

\url{}
\footlineauthor{}


\maketitle

\begin{article}

\begin{abstract} Many studies have shown that a mechanical displacement of the axonal membrane accompanies the electrical pulse defining the Action Potential (AP).  Despite a large and diverse body of experimental evidence, there is no theoretical consensus either for the physical basis of this mechanical wave nor its interdependence with the electrical signal.  
In this manuscript we present a model for these mechanical displacements as arising from the driving of surface wave modes in which potential energy is stored in elastic properties of the neuronal membrane and cytoskeleton while kinetic energy is carried by the axoplasmic fluid.  In our model these surface waves are driven by the traveling wave of electrical depolarization that characterizes the AP, altering the compressive electrostatic forces across the membrane as it passes.  This driving leads to co-propagating mechanical displacements, which we term Action Waves (AWs). Our model for these AWs allows us to predict, in terms of elastic constants, axon radius and axoplasmic density and viscosity, the shape of the AW that should accompany any traveling wave of voltage, including the AP predicted by the Hodgkin and Huxley (HH) equations.  We show that our model makes predictions that are in agreement with results in experimental systems including the garfish olfactory nerve and the squid giant axon.  We expect our model to serve as a framework for understanding the physical origins and possible functional roles of these AWs in neurobiology.
\end{abstract}


\abbreviations{HH Hodgkin Huxley AP Action Potential AW Action Wave}

\section{Significance}
Neurons communicate using Action Potentials (APs) that propagate along axons.  While most studies have focused on the AP's electrical signatures, many studies have reported the existence of a mechanical wave traveling alongside the electrical pulse.  Here we hypothesize that these mechanical waves, which we term Action Waves (AWs), are driven by the electrical changes that accompany the AP, just as a speaker uses charge separation to drive sound waves.  We argue that these driven displacements are surface waves involving both the axonal membrane and its surrounding fluid. In addition to providing a theoretical framework explaining many existing experiments we expect that a better understanding of the physical basis for AWs will shed light on their possible biological significance.

\section{Introduction}
For many decades, Action Potentials (APs) have been measured using electrophysiological methods and understood as electrical signals generated and propagating along the axonal membrane~\cite{Hodgkin52}. While measuring and understanding the electrical component of the AP has been the focus of most experimental and theoretical efforts, a large number of experimental studies have shown that the AP is accompanied by fast and temporary mechanical changes. These include changes in axonal radius ~\cite{Cohen73,Hill77,Iwasa80,Tasaki88a,Tasaki88b,Tasaki89,Tasaki90,Tasaki92,Tasaki82b}, pressure~\cite{Iwasa80,Barry70}, optical properties~\cite{Tasaki93}, the release and subsequent absorption of a small amount of heat~\cite{Tasaki89} and shortening of the axon at its terminus when the AP arrives~\cite{Tasaki82}. 
 
Despite this wealth of experimental evidence, the physical basis for the mechanical and thermal signals that accompany the AP remains poorly understood. 
To our knowledge, there have been no attempts to quantitatively describe the mechanical component of the AP as an electrically driven phenomenon.  In this manuscript we consider a minimal mechanical model of the axon as an elastic and dielectric tube filled and surrounded with viscous fluid.  We show that as the AP passes, changes in charge separation across the dielectric membrane alter surface forces that act on the membrane's geometry.  As we show, these forces lead to co-propagating displacements which we call Action Waves (AWs).


The electrically driven mechanical modes which we consider are surface waves in which potential energy is stored in the elastic energy associated with deforming the axonal surface, while kinetic energy is carried by the axoplasmic and extracellular fluid, primarily along the direction of propagation.  The physics of these modes are governed by the viscosity and density of the axoplasm, the radius of the axon, and the elastic moduli of the axonal membrane and cytoskeleton.
In this manuscript we estimate the displacement and axoplasmic flow field that accompanies an AP in an axon and compare our results with those seen in two experimental systems: the squid giant axon and the garfish olfactory nerve. We also briefly address the issue of thermal changes associated with action potential propagation.


Previous theoretical efforts have considered mechanical modes independent from the electrical AP.   Some have considered a nonlinear solitonic excitation~\cite{Heimburg05,Mosgaard13} in which a compression wave in the membrane travels at its speed of sound, with a portion of the membrane passing through a reversible freezing transition.  Others have considered propagating axoplasmic pressure pulses driven by actomyosin contractility~\cite{Rvachev10}.
Our results suggest that it is possible to integrate existing experiments on mechanical aspects of the AP into our understanding of nerve pulse propagation.  We hope that this framework for the physical underpinnings of the AW will help focus future experiments towards understanding their possible functional roles.


 \begin{figure*}
\centering
\caption{\label{fig:fig3D} (A) The Action Potential and accompanying Action Wave constitute an electromechanical pulse traveling along the axon.  The initially depolarized axonal membrane (left, orange $+$s and $-$s) is depolarized as the AP passes (center).  This leads to changes in the electrostatic forces acting on the membrane (grey tube), which drives a co-propagating mechanical surface wave.  This surface wave includes a deformation of the axonal surface as well as a displacement of the axoplasmic and extracellular fluid ( $\vec{\Delta}$ green arrows).  In our model, kinetic energy is carried in the motion of the 3D fluid and is given by $\mathcal{T}=\int 2\pi \rho d\rho dz \frac{\rho_{3D}}{2} (\frac{\partial \vec{\Delta}(\rho,z)}{\partial t})^2$ where $\rho_{3D}$ is the density of water.  (B) Potential energy is stored in deformation of the axonal surface.  $h$ defines a relative height field, and $l$ a lateral stretch field, which describe deviations from an unstretched tube of radius $r_0$ as shown in the figure and described in the results.  In terms of these fields the potential energy is given by $\mathcal{U}=\int dz (\pi r_0 \kappa h^2 (z) -F^h(z)h(z)-F^{l}l(z))$ where $F(z)$s are electrostatic forces estimated in Results and $\kappa$ is an elastic constant of the axonal surface.  
 }
\includegraphics[width=\textwidth]{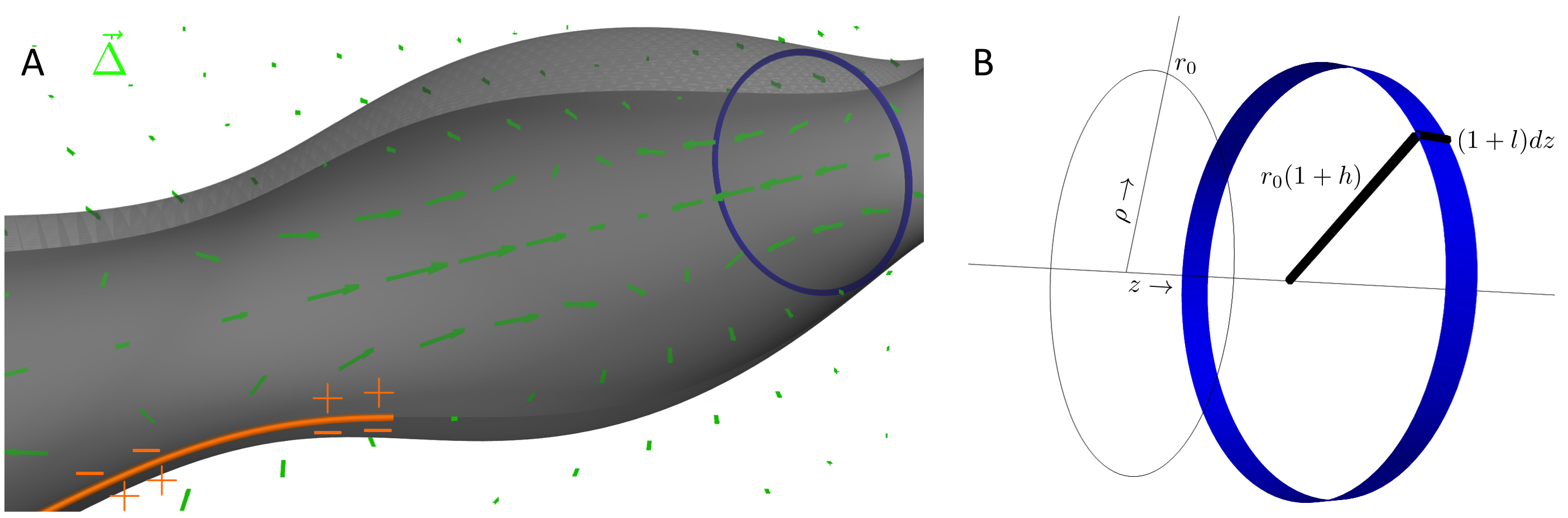}
\end{figure*}

\section{Results}
We first define the model shown schematically in Fig.\ref{fig:fig3D} and calculate the susceptibility of displacement fields to time dependent driving at the axonal surface.  We then estimate the driving force on these modes that accompanies the passing of an AP.  Together these allow us to estimate the AP's co-propagating displacement field.  We then estimate the thermal signatures of the electromechanical wave predicted by our model.  Finally we compare our predictions to results from several experimental systems.  
\subsection{Axons support mechanical surface modes} 

Axially symmetric flows are described by a time dependent displacement field $\vec{\Delta}(\rho,z,t)=\left\{ \Delta_\rho(\rho,z,t), \Delta_z(\rho,z,t) \right\}$ (green arrows in Fig.\ref{fig:fig3D}A).   We parametrize these using an orthogonal basis, $\tilde{\Delta}_{k,q}$, for the space of axially symmetric and incompressible flows that diagonalizes the vector Laplacian (see supplement):


\begin{equation}
\vec{\Delta}(\rho,z)= \int_{0}^{\infty} \frac{dq}{\pi} \int_{-\infty}^{\infty} \frac{dk}{2\pi}  e^{ikz} \left\{ \frac{k}{q} J_1(q \rho), iJ_0(q \rho) \right\}\tilde{\Delta}_{k,q} 
\end{equation}
where $J_0(x)$ and $J_1(x)$ are Bessel functions of the first kind. 

We consider the axon to be a tube which, in the absence of displacement, has radius $r_0$ and which is infinite in extent in the $z$ direction\footnote{Many neurons are in a regime where this simplification is reasonable- for example the garfish nerve is approximately $20cm$ long, and the width of its action potential is $\sim 2mm$, roughly $100 \times$ smaller.}.  
We assume that the displacement field in the bulk carries the axonal surface along with it, through non-slip boundary conditions.  Distortions of this surface can be described by two fields that are entirely determined by $\Delta$: a relative height field $h(z,t)=\Delta_\rho(r_0,z,t)/r_0$ and a lateral stretch field $l(z,t)=\frac{\partial}{\partial z}\Delta_z(r_0,z,t)$ (see Fig.\ref{fig:fig3D}B)). These can be expressed in Fourier space as: 
\begin{equation}
\label{eq:sdef}
\tilde{h}_k(t) =k \int_{0}^{\infty} \frac{dq}{\pi}\frac{1}{qr_0}J_1(q r_0) \tilde{\Delta}_{k,q}(t) 
\end{equation}
 In terms of these fields we can write the kinetic energy $\mathcal{T}$, and the potential energy $\mathcal{U}$ as:
\begin{equation}
\label{eq:UTdef}
\begin{array}{rl}
\mathcal{T} &=\frac{\rho_{3D}}{2}\int_{0}^{\infty} \frac{dq}{\pi}  \int_{-\infty}^{\infty} \frac{dk}{2\pi} \frac{-2 \left(q^2+k^2\right)}{q^3} \frac{\partial \tilde{\Delta}_{k,q}}{\partial t}  \frac{\partial \tilde{\Delta}_{-k,q}}{\partial t} \\ \\

\mathcal{U} &=  \int_{-\infty}^{\infty}  \frac{dk}{2\pi} \frac{2\pi r_0 \kappa}{2}\tilde{h}_k \tilde{h}_{-k} -\tilde{h}_k \tilde{F}^h_{-k}-\tilde{l}_k \tilde{F}^l_{-k}

\end{array}
\end{equation}
Where $\rho_{3D}$ is the density of the bulk fluid, $\kappa$ is a 2D compressibility modulus of the axonal surface and where $F^{h/l}$ are externally applied fields.  Equations of motion for the bulk field are given by the linearized Navier Stokes equations with each degree of freedom $\tilde{\Delta}_{k,q}$ feeling an additional force given by $-\frac{\partial \mathcal{U}}{\partial \tilde{\Delta}_{k,q}}$, 
where the potential energy $\mathcal{U}$ depends on the $\Delta$s through their dependence on $h$. After Fourier transforming in time, these equations can be combined to yield equations for the $\Delta$s in terms of driving forces $F$ and the as yet to be solved for field $h$ (see supplement):
 \begin{equation}
 \label{eq:EOMDEL}
\Delta_{k,q,\omega}=\frac{\frac{k}{r_0} J_1(qr_0)\left(2\pi r_0 \kappa \tilde{h}_{k,\omega} - \tilde{F}^h_{k,\omega}\right)-kqJ_0(qr_0)\tilde{F}^l_{k,\omega}}{2\rho_{3D} \omega^2+i\omega (\eta/\rho_{3D})(q^2+k^2)} 
 \end{equation}
 

Integrating the left hand side of eq.\ref{eq:EOMDEL} according to eq.\ref{eq:sdef} 
allows us to remove the $\Delta$s from the equation and solve for $\tilde{h}_{k,\omega}$ which are linear in the driving fields and can thus be written as $\tilde{h}_{k,\omega}=\chi^{hh}_{k,\omega}F^{h}_{k,\omega}+\chi^{hl}_{k,\omega}F^{l}_{k,\omega}$
where  $\chi$s are  susceptibilities.  Collecting terms and taking the limit relevant to action potentials where $k r_0 >> 1$ we can write:
\begin{equation}
(\chi^{hh}_{k,\omega})^{-1} = 2\pi r_0 \kappa-\frac{2\pi \rho_{3D} r_0^2 \omega^2}{k^2M_{11}(\alpha)} 
\end{equation}
where $\alpha=\frac{\rho_{3D} r_0^2 \omega}{\eta}$ is a Reynolds-like number and where $M_{11}(\alpha)$ is given by:
\begin{equation}
\label{eq:defM}
M_{11}(\alpha) =\int_{0}^{\infty}  \frac{dx}{x}\frac{J_1(x)^2}{1+i x^2/\alpha} 
\end{equation}
Notice that this solution for $h$, in conjunction with eq.\ref{eq:EOMDEL} (and a similar form for $\chi^{hl}$, see supplement) allows us to solve for arbitrary $\Delta$s, and therefore arbitrary displacement fields induced by driving at the axonal surface. 

The dispersion relationship for these waves is given by $k(\omega)=\omega \sqrt{r_0\rho_{3D}/\kappa  M_{11}(\alpha)}$.   From this we can define the propagation velocity $c_{pr}(\omega)=\omega/\mathcal{R}(k(\omega))$~\cite{Griesbauer12}.  When $\alpha<<1$ as is relevant to most APs in non-myelinated neurons, $c_{pr} \sim \sqrt{\kappa r_o \omega /\eta}$.  When $\alpha >>1$ as is relevant to the squid giant axon, $c_{pr}\sim \sqrt{\kappa/\rho_{3D} r_0}$.  We emphasize that the driven waves we consider will travel at the speed of the electrical AP that drives them, $c_{AP}$, and not at their undriven propagation speed $c_{pr}$.  However, if these speeds are closely matched then interesting resonant phenomenon occur as discussed below.  Interestingly, most neurons are in the regime where $c_{pr} \sim r^{1/2}$ as predicted by the cable theory and observed experimentally. 





\subsection{Mechanical modes are driven by changes in charge separation across the membrane}

\begin{figure}
\centering
\caption{\label{fig:diffC} (A) We consider the electrical component of the AP to be a Gaussian wave of depolarization, and take the asymmetry of the surface potential $\Delta \psi$ (orange) to be equal to the resting potential.  (B) This traveling wave of depolarization leads to a traveling force on the mechanical modes as described in the text. This force leads to a mechanical response which travels alongside the AP as described in the text (C-F). Two components of this wave, the radial displacement of the membrane (C) and the average longitudinal displacement inside of the axon (D) are plotted in the parameter regime where $\alpha <<1$ and for $c_{pr}>c_{AP}$.  Ahead of the AP there is a longitudinal displacement backwards relative to the direction of propagation and a slightly inward movement of the membrane.  The depolarization of the membrane is accompanied by a swelling and a lowering of the displacement field.  After the AP passes, the mechanical modes quickly relax.  (E, F) The response to the same driving is shown in the regime where $\alpha<<1$ and $c_{pr}<c_{AP}$. }
\includegraphics[width=3.00in]{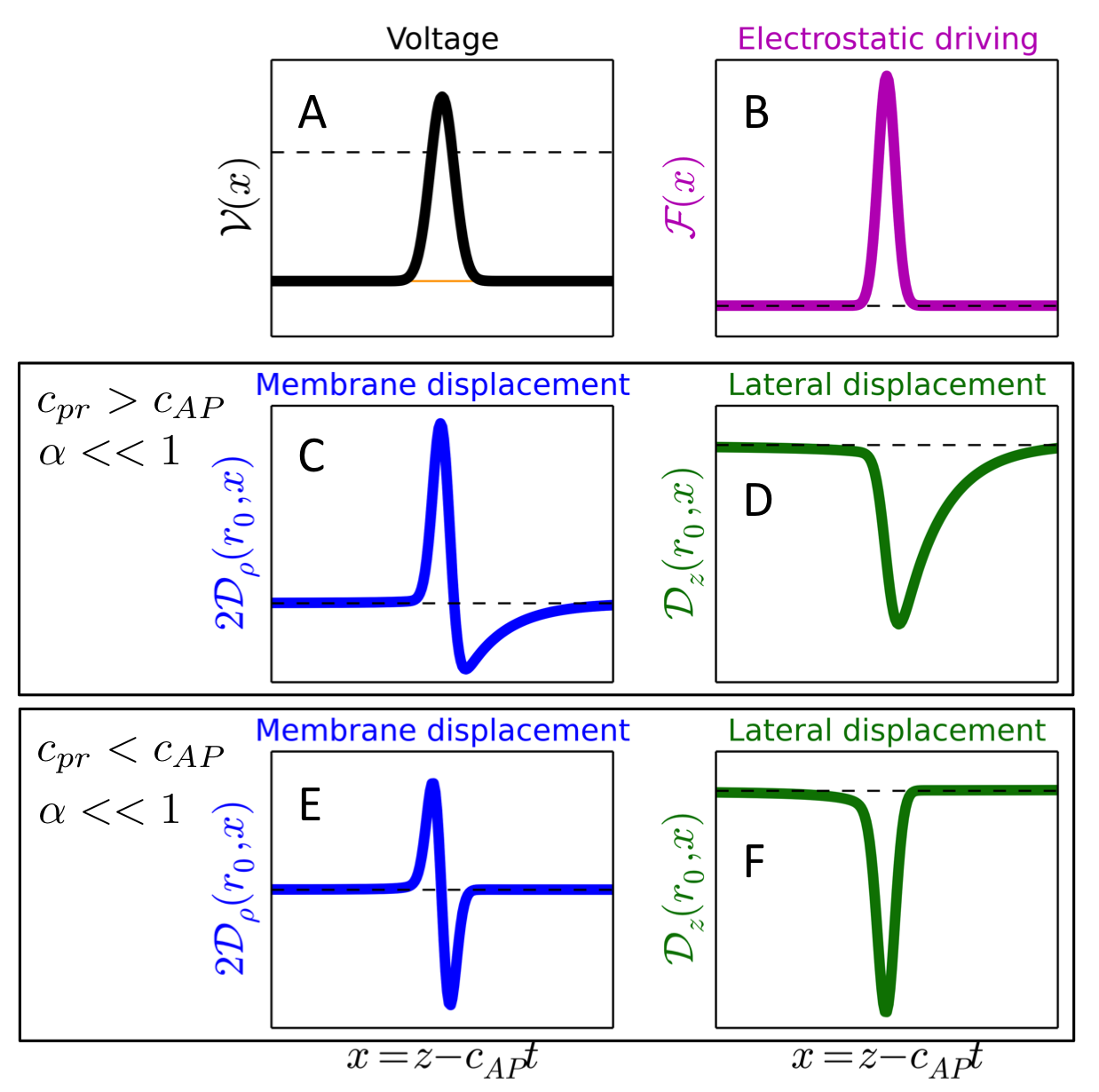}
\end{figure}

  We consider the electrical component of the AP to be a traveling wave of voltage, taking the form $V(z,t)=\mathcal{V}(x)$, where $x=z-c_{AP}t$ is a co-moving coordinate and $c_{AP}$ is the propagation speed of the action potential.  Here we take a gaussian form for $\mathcal{V}(x)$ as plotted in Fig.\ref{fig:diffC}A.  The axonal membrane is composed of charged elements which leads to a surface potential $\psi_{in/out}$ on the inner and outer leaflets~\cite{Alvarez78}.  As axonal membranes are not symmetrically charged, the potential difference across the membrane is given by $V_m(z,t)=V(z,t)-\Delta \psi$ where $\Delta \psi=\psi_{in}-\psi_{out}$ is the difference in surface potentials between the inner and outer leaflets.  Throughout, we consider $\Delta\psi$ equal to the resting potential for simplicity- see Fig.\ref{fig:diffC}A, as suggested by the few experiments where it has been measured~\cite{Alvarez78}.  As $V_m$ changes, the compressive forces on the membrane are altered leading to changes in shape (electrostriction) and corresponding changes in capacitance~\cite{Alvarez78}.  

The charge density at the membrane is given by $q_m(r)=C_0 V_m(z)$, where $C_0$ is the capacitance per unit area of the unstretched membrane.  Here we allow the shape of the membrane to vary at fixed charge per unit membrane.  To calculate the resulting forces on the membrane's geometry we use that $r(z)=r_0(1+h(z))$ while the length of an infinitesimal cylinder of membrane is given by $(1+l(z))dz$ (see Fig.\ref{fig:fig3D}B).  We can estimate the capacitance of this infinitesimal cylinder of membrane $C(z)dz$ by taking the membrane to be incompressible in $3D$, and by further assuming that it is an isotropic material so that its capacitance per unit area is inversely proportional to the membrane thickness\footnote{this approximation likely underestimates the effect of electrostriction~\cite{Heimburg12}.}.  In this approximation $C(z)dz =2\pi (1+h(z))r_0 (1+l(z))dz C_0(1+h(z)+l(z))$.
The capacitive energy stored in the membrane is thus given by:
\begin{equation}
\label{eq:defUc}
\mathcal{U}_c = \int dz \frac{q_m^2(z) }{2 C(z)} 
\approx \int dz \pi r_0 C_0 V^2_m(z)\left(1-2 h(z) -2l(z) \right) 
\end{equation}

Including this in the potential energy defined in eq.\ref{eq:UTdef} we see that $F^h(z,t)=F^l(z,t)=2\pi r_0 C_0 V_m^2(z,t)$.  Specializing to a traveling wave of voltage, $V(z,t)=\mathcal{V}(x)$, $F^{h}(z,t)=\mathcal{F}(x)=2\pi r_0 C_0 \mathcal{V}_m^2(x)$. 
This driving force is plotted in Fig.\ref{fig:diffC}B.

This driving only activates modes for which $\omega=c_{AP} k$, and therefore the action potential is accompanied by a co-moving displacement field, $\vec{\Delta}(\rho,z,t)=\vec{\mathcal{D}}(\rho,x)$, where again $x=z-c_{AP}t$.  Although the mechanical response always travels in a co-moving frame with the electrical component driving it, many of its properties are determined by whether the mechanical propagation speed $c_{pr}$ is faster, slower, or comparable to $c_{AP}$ at relevant wavelengths.  In Fig.\ref{fig:diffC} we show two components of the displacement field.  First, the membrane displacement, $2 \mathcal{D}_\rho(r_0,x)$ (Fig.\ref{fig:diffC}c,e), as observed in experiments which measure axonal swelling, and secondly the average lateral displacement inside of the axon, $\bar{\mathcal{D}}_z(\rho<r_0,x)=1/\pi r_0^2\int_0^{r_o}2\pi \rho d\rho \mathcal{D}_z(\rho,x)$ (Fig.\ref{fig:diffC}d,f).  These are plotted vs. $x$ for the case where $\alpha <<1$ 
both for $c_{pr}>c_{AP}$ (Fig.\ref{fig:diffC}c,d) and for $c_{pr}<c_{AP}$ (Fig.\ref{fig:diffC}e,f).

\subsection{Thermal effects arise from the co-propagating electromechanical soliton}

Previous authors have investigated heat production during AP propagation experimentally~\cite{Tasaki89,Howarth75,Howarth79a,Howarth79b} in the garfish olfactory nerve, finding the APÕs depolarization associated with a small release of heat, and its repolarization associated with absorption of heat of comparable magnitude.  We consider two sources of heating that we expect to accompany AP propagation.  Firstly, the HH equations already predict heat production with a similar qualitative shape to that seen in experiments~\cite{Nogueira83}.  The sodium influx associated with depolarization is exothermic.   Na$^+$ ions travel down their concentration and potential gradients so that as the membrane depolarizes the capacitive energy stored in the membrane is released as heat.  However, the efflux of K$^+$ ions is endothermic, with K$^+$ ions traveling up their potential gradients but down their concentration gradients, converting heat in the environment into capacitive energy at the membrane. Thus the HH model predicts that in total, no heat is produced as the AP passes- APs are driven entirely by the entropy of mixing.  At the peak of the action potential, given a membrane capacitance of $1\times10^{-6}F/cm^2$~\cite{Gentet00}, the discharging of  $70mV$ associated with the AP is expected to release approximately $\mathcal{Q}_{e}=.25 \mu J/m^2$ of electrical energy as heat.

Our model predicts an additional contribution to the heat production associated with AP propagation that arises from the approximately isothermal\footnote{This is not a sound wave, and the membrane's intrinsic time constant is much faster than the $ms$ time-scales associated with the AP, suggesting the process is nearly isothermal, as assumed here.  This assumption could be called into question if the membrane is near to a phase transition~\cite{Veatch08} where its equilibration might acquire a hierarchy of longer time scales.} distortion of the membrane as the AW passes.  In the absence of driving, the membrane's geometry locally minimizes the free energy so that $\frac{\partial{F}}{\partial A}=\frac{\partial E}{\partial A}-T \frac{\partial S}{\partial A}=0$.  In equilibrium, an entropic force stretching the membrane balances an energetic force pulling the membrane in, each of magnitude $T \alpha\kappa$, where $\alpha$ is the thermal expansion coefficient of the membrane defined by $dA/A= \alpha dT$.  Thus, our theory predicts an additional release of mechanical energy as heat $\mathcal{Q}_{m} = -T \alpha \kappa \left (h(z)+l(z) \right)$.  We ignore two much smaller sources of heating: viscous heating from the flow of axoplasm itself, and a coupling of our height field to $V^2$ through the changing capacitance. 

\subsection{The model predicts mechanical and thermal responses that are in quantitative agreement with existing experimental results}
\begin{figure*}
\centering
\caption{\label{fig:GarandSquid} Quantitative comparison with three experimentally favored systems using model and parameters as described in the text.  \textit{Voltage:} (A,G,J) For each we take as input the shape of the electrical component of the AP as well as its speed from the experimental literature.  \textit{Displacement:} Using these in conjunction with our model we make predictions for the membrane displacements (B,H,K) and lateral displacements of the axoplasm(C,I,L).  \textit{Heat: }  For the Garfish Olfactory nerve we also consider thermal effects. We plot the electrostatic heat (D) released by the depolarization associated with the AP, the mechanical heat (E) released by the distortion associated with the AW as well as a prediction for the experimentally observable total heating (F). }
\includegraphics[width=\textwidth]{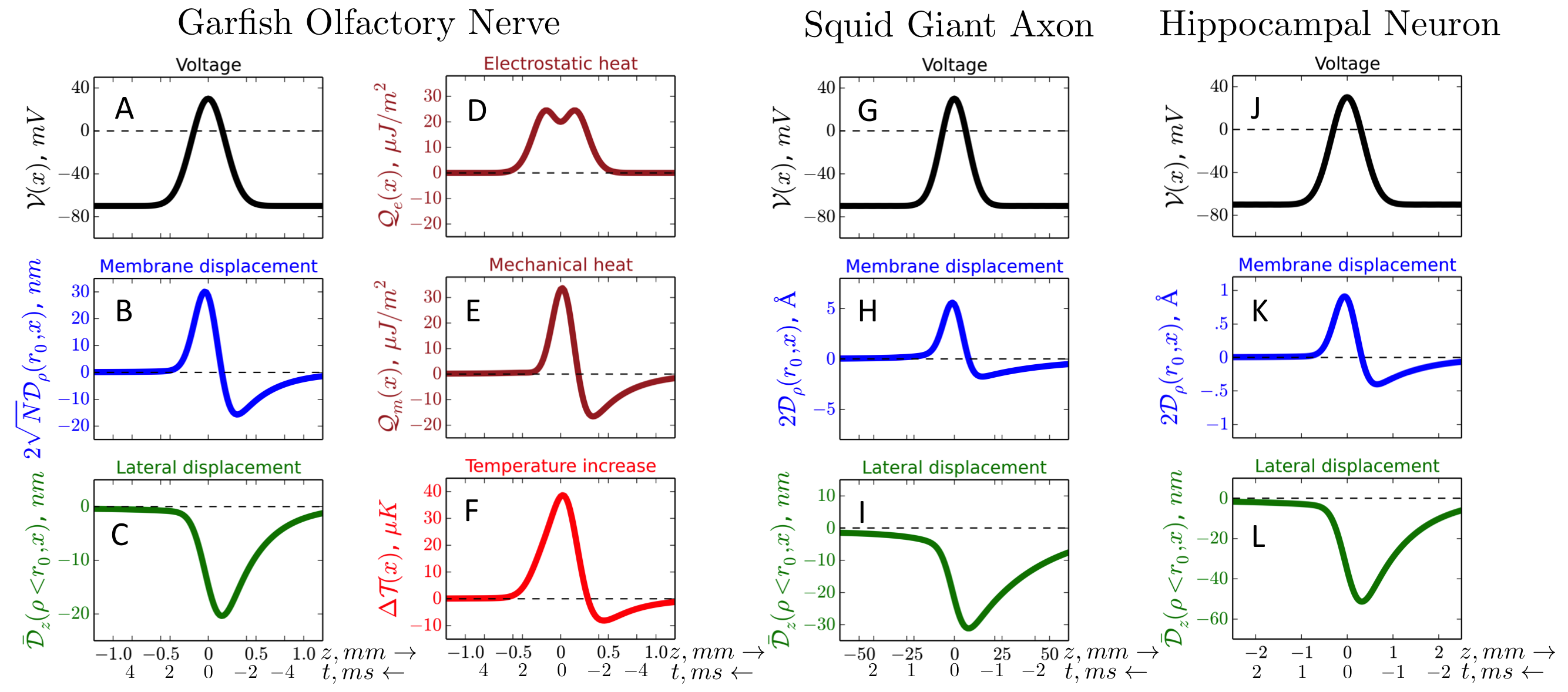}
\end{figure*}

We first consider the squid giant axon whose mechanical response has been considered experimentally in~\cite{Tasaki82b,Terakawa83,Tasaki92}. To quantitatively compare to these experimental results, we use our model described above with parameters from~\cite{Tasaki82b} for the squid's AP speed, $c_{AP}=21.2 m/s$, AP width of $20mm$ (fig~\ref{fig:GarandSquid}A)  and axonal radius $r_0 = 238 \mu m$.  Although the mechanical properties of the squid giant axon's sheath and cytoskeleton are not known quantitatively enough for our purposes, we use $\kappa=300J/m^2$ for the relevant stretch modulus, which is in the regime expected for the $10 \mu m$ sheath of connective tissue surrounding the squid giant axon~\cite{Brown90} presuming the sheath has a 3D Young's modulus around $1GPa$~\cite{Broedersz14}. This choice puts our model into the regime where $c_{pr}>c_{AP}$.  In all of our results we take the viscosity of axoplasm $\eta$ to be three times the viscosity of water~\cite{Haak76} and assume that the density of the medium $\rho_{3D}=10^3kg/m^3$ is that of water. Our model's prediction for the membrane's displacement are shown in fig~\ref{fig:GarandSquid}H in agreement with both the size and shape of the mechanical response seen in~\cite{Tasaki82b}.  We also plot the average lateral displacement in fig~\ref{fig:GarandSquid}I which is a novel prediction of our model which has not been directly measured.  

We next consider the garfish olfactory nerve bundle, which is around $1mm$ in diameter and contains approximately $N=10^7$ non-myelinated axons each of radius $r_o \approx .125 \mu m$ and with total length of $20cm$~\cite{Easton71}.  We use the speed of AP propagation to be $c_{AP}=.2m/s$ and the width of the AP to be $.4mm$~\cite{Tasaki90} (fig~\ref{fig:GarandSquid} A).  We use $\kappa=1J/m^2$,  in the range suggested by results showing periodic actin rings with periodicity of $\sim 180 nm$~\cite{Xu13,Broedersz14}.  This is chosen primarily so as to stay in the regime where $c_{pr}>c_{AP}$.  To compare with experimental results in~\cite{Tasaki90} we assume that each axon swells independently, so that the observed response is $2\sqrt{N}\mathcal{D}_\rho$.  This total response of the nerve is plotted in fig~\ref{fig:GarandSquid}B, which can be compared to experimental results in ~\cite{Tasaki90,Tasaki89}. We also predict a lateral displacement in the axon (fig~\ref{fig:GarandSquid}C), which has not been measured, although a  shortening of the axon of comparable magnitude, roughly coincident with AP arrival has been seen in Crab motor neurons~\cite{Tasaki82}.  

We also compare model predictions to thermal measurements made in the garfish olfactory nerve where the $10^7$ axons contain roughly $6\times 10^4cm^2$ of membrane per $cm^3$ of volume~\cite{Easton71}.  As described in the previous section, the HH model already predicts that the activation of all of these nerves at once will lead to the release of electrical energy as heat as plotted in fig~\ref{fig:GarandSquid}D.  Using $T \alpha \kappa=1J/m^2$, our model predicts that the AW releases an additional and similarly sized amount of mechanical energy as heat as shown in fig ~\ref{fig:GarandSquid}E.  Assuming that all of the axons are activated simultaneously and that this heat is absorbed by a bulk with heat capacity close to that of bulk water, we predict that a thermometer would measure a small heating and subsequent cooling as shown in fig~\ref{fig:GarandSquid}F.

Finally we make predictions for unmyelinated rat hippocampal neurons using $r_0= 1\mu m$, $c_{AP}=1m/s$, and $\kappa=1J/m^2$.  Our results plotted in fig~\ref{fig:GarandSquid}J-L are new predictions of our model that could be tested in the near future.

\section{Discussion}
Here we have presented a biophysical model for a mechanical AW that accompanies the electrical component of the AP.  In our model, electrical driving excites surface waves in which potential energy is stored at the axonal surface and kinetic energy is carried by the bulk fluid.  While the mechanical properties of these waves depend on many biophysical parameters discussed below, our model is already in agreement with a range of experimental results~\cite{Tasaki88a,Tasaki88b,Tasaki89,Tasaki90,Tasaki92,Tasaki82}.


The vast majority of theoretical and experimental efforts in neuroscience have been aimed at understanding the electrical component of the AP, using the HH framework. The primary focus of this paper is orthogonal: to understand the mechanical component as driven by the well characterized electrical component of the AP.  Our model does not require an underlying theory of how this electrical component arises.  We emphasize that any traveling electrical wave will induce a co-propagating mechanical wave, whether arising from an HH-like underlying mechanism, or one involving feedback form the mechanical wave itself.   

Our model differs from most other models of the mechanical response existing in the literature in that it is electrically driven, by the depolarization wave that forms the AP.   Where driving has been explicitly considered it has been taken to arise from actomyosin contractility~\cite{Rvachev10}.  While we cannot rule out a possible role for actomyosin contractility as is certainly relevant in cardiomyocytes~\cite{Bers02}, the magnitude of the experimentally observed mechanical response does not require extra driving of this form.  

Our model also differs in the nature of the modes involved.  Previous authors~\cite{Heimburg05, Andersen09} have considered a nonlinear sound wave at the peak of which the membrane has undergone a 2D freezing transition. These authors assume that this solitonic wave independently propagates without dissipation and therefore does not require a continuous input of energy from electrical driving.  In our model, only potential energy is stored at the axonal surface, while kinetic energy is carried by the bulk fluid.  In \cite{Heimburg05, Andersen09} both kinetic and potential energy are stored in the membrane itself.  Thus displacements in the axoplasm are a prediction of our model distinct from that in \cite{Heimburg05, Andersen09}.  Our model assumes that, since these displacements are relatively small, we can assume that modes are in the linear response regime.  This is in contrast to \cite{Heimburg05, Andersen09, Heimburg12} in which mechanical nonlinearities, rather than electrical ones are responsible for the solitonic shape of the AP.  

Our model makes very different predictions for the thermal signatures that accompany the AP.  One prediction of \cite{Heimburg05, Andersen09} is that the AP should be accompanied by a very large thermal response arising from the latent heat of melting.  We predict a much smaller heat signature arising from two distinct mechanisms; The release and subsequent absorption of capacitve energy as heat, and mechanical heat from the stretching of the membrane in the almost isothermal regime.  This smaller heat signature is a key prediction distinguishing our model from~\cite{Heimburg05, Andersen09}.  

The mechanical modes we consider are similar to those studied experimentally in~\cite{Shrivastava13} who look at the propagation of surface waves in monolayers at the air-water interface.  While these experiments provide inspiration for our model, there are important differences.  In addition to using monolayers rather than bilayers, the planar geometry used in~\cite{Shrivastava13} leads to qualitatively different dispersion relations and mechanical properties. 

Much of the uncertainty in our model arises from our uncertain knowledge of the elastic properties of the axon which are determined by a combination of the elastic properties of the axonal membrane and the architecture of the axonal cytoskeleton, each being the focus of intense inquiry~\cite{Conde09,Xu13,Fletcher10,Lukinavicius14}.  
The axoplasmic membrane and associated cytoskeletal fibers make up the axonal surface which contributes to the elastic energy of deformations through surface terms. These enter our model in terms of the 2D displacement fields $h$ and $l$.  The membrane contributes a term which punishes stretching through elastic modulus $\kappa_{2D}$ which multiplies $\int dz \pi r_0(h(z)+l(z))^2$ in the potential energy $\mathcal{U}$.  It also contributes a term that punishes changes in total area through tension $\sigma$ which multiplies $\int dz \pi r_0^3(\frac{\partial h }{\partial z})^2$ in $\mathcal{U}$. This term is subdominant in the regime where $kr_0<<1$, and so we drop it here.  Recent experiments have demonstrated that the axonal membrane cytoskeleton is exquisitely ordered, forming a quasi-1D lattice, with evenly spaced actin rings~\cite{Xu13}. These likely contribute a term to the potential energy that depends only on the height field $h$, motivating our choice of potential energy in eq.\ref{eq:UTdef}.  The same study found longitudinal spectrin fibers, which likely contribute a term to the potential energy proportional to $\int dz \pi r_0 l^2(z)$.

The bulk of the axon contains microtubules~\cite{Fletcher10} which are arranged parallel with the axons. The present analysis focuses on the elastic properties of the surface, assuming that the bulk is purely viscous, with little elastic resistance to shear.  This is motivated by the parallel microtubule arrangement within the axon, which should lead to a very anisotropic shear modulus.   In our model the dominant shears are orthogonal to the energetically unfavorable directions.  Thus, while it is possible that the bulk elasticity makes important contributions to mechanical properties relevant for the AW, we ignore them here.

Our finding that mechanical AWs of the size and shape seen experimentally naturally accompany the electrical AP suggests the possibility that AWs might exist even without functional significance.  On the other hand, it seems likely that biology would take advantage of this co-propagating information, especially if real neurons are indeed in the regime where $c_{pr}>c_{AP}$ in which the mechanical component arrives before the electrical AP.  We discuss several mechanistic possibilities.  

AWs could feed back and influence the electrical component of the AP.  Such a picture is broadly supported by experiments that see changes in electrical properties and even the triggering of APs on mechanical stimulation of single neurons~\cite{Terakawa82,Julian62,Galbraith93}.  
Furthermore, particular cell types are known to have important electro-mechanical coupling, most notably cardiac cells~\cite{Quinn12} and stereocilia of the inner ear~\cite{Reichenbach14}.   Intriguingly, these stereocilia sense excitations of similar surface waves propagating in the cochlea, albeit in a very different geometry. Mechanics could influence electrical properties by direct alteration of membrane conductance and/or capacitance or by influencing the gating of voltage gated ion channels.  Studies in artificial lipid bilayers without protein channels have shown ion channel like conductances sensitive to stretching, although their existence in neuronal membranes and relevance to the AP has not been established \cite{Antonov80, Mosgaard13, Blicher13}. 

Membrane mechanical properties are also known to influence some ion channels directly~\cite{Anishkin14} 
including the voltage gated sodium channels responsible for the AP~\cite{Tabarean02,Beyder10}. 
We consider two ways in which channels could acquire mechanical sensitivity.  First, a channel could sense membrane tension by changing its 2D membrane footprint by  $\Delta A$ upon activation.  Such a channel would have its gating altered substantially by changes in tension greater than $k_B T/\Delta A$~\cite{Anishkin14}.  The MscL channel, a bacterial osmotic pressure sensor for which $\Delta A \sim 20nm$, can detect changes in tension smaller than $10^{-3}N/m$~\cite{Anishkin14}, while measured mechanosensitivity of Na channels is smaller, with a change in tension closer to $10^{-2}N/m$ required~\cite{Beyder10}.  Our model predicts changes in tension of order $\kappa_{2D} (l+h) \sim 10^{-4}$, close to the limit of sensitivity for dedicated tension sensors, but likely irrelevant for Na channel function.  More plausibly relevant for the AP, a channel could directly sense displacements by connecting to a cytoskeletal element through a tether whose length is coupled to channel activity~\cite{Mueller14}. Such a mechanism is thought to convert mechanical signals in the inner ear into electrical ones, where channels can sense displacements of $5\mu m$ hair cells by as little as $0.3nm$, about as large as the radial displacements predicted by our model for typical neurons, and much smaller than the horizontal ones.  Though this mechanism remains speculative even in the inner ear, Na channels do coordinate at nanometric scales with periodic actin rings and spectrin~\cite{Xu13}.  Na channels are anchored to the cytoskeleton by ankyrin-G, which is homologous to the ankyrin repeats in specialized TRP channels widely believed to play the role of the mechanical to electrical transducer in the inner ear~\cite{Corey04}.

Our model adds to a more complete theory for action potential generation and propagation that explains electrical, mechanical, chemical and thermal aspects. We hope this more complete theoretical understanding of the phenomena accompanying the AP will guide experiments seeking to understand the role of the striking cytoskeletal ultrastructure recently available to state of the art super-resolution techniques~\cite{Xu13,Lukinavicius14}.  We foresee future theoretical and experimental work will highlight the extent to which AWs play a functional role in neuronal information processing.




\begin{acknowledgments}
The authors would like to thank Pepe Alcami, Francisco Balzarotti, William Bialek, Cliff Brangwynne, Chase Broedersz, Christopher Battle, Elisa D'Este,  Federico Faraci, Thomas Heimburg, Stefan Hell, Andrew Leifer, Georg Rieckh, James Sethna, Joshua  Shaevitz, Howard Stone, Sarah Veatch, Ned Wingreen for fruitful discussions and their enthusiasm for our ideas. Both authors are grateful to ICTP in Trieste, and BM thanks a Lewis Sigler fellowship for support.
\end{acknowledgments}

\newpage
\noindent \textbf{\LARGE Supplementary Text}



\section{Orthogonality and Mode Decomposition}
\label{sec:ortho}
In this section we find an appropriate basis for displacement modes in 3D.    We first seek a complete and orthogonal basis for a displacement field, $\vec{\Delta}(\rho,z)=\left\{ \Delta_\rho(\rho,z), \Delta_z(\rho,z) \right\}$ which is divergence free, cylindrically symmetric, and which diagonalizes the vector Laplacian.   To capture the flow field both inside and outside of the axonal membrane, we embed the system inside a larger cylinder of radius $R$ and length $L$ both of which we will take to be infinite.  At $R$ we take BCs in which  $\Delta_z(R,z)=0$, chosen for convenience and arbitrary as $R \rightarrow \infty$. We similarly take our system to be periodic in the z-direction.  We construct a general radially symmetric displacement field out of the two parameter $(k,q)$ basis fields: 
\begin{equation}
\vec{\Delta}(\rho,z)=\left\{ \Delta_\rho(\rho,z), \Delta_z(\rho,z) \right\}=\left\{ \frac{k}{q}e^{i k z}J_1(q\rho), i e^{i k z}J_0(q\rho) \right\}
\end{equation}
Where $J_0(x)$ and $J_1(x)$ are Bessel Functions of the first kind.  Using the Bessel function identities 
\begin{equation}
\begin{array} {rl}
\frac{\partial}{\partial x} \left[ x^m J_m(x) \right] &= x^m J_{m-1} \\
J_{-m} &=(-1)^m J_m
\end{array}
\end{equation}
 it can be verified that these flow fields are incompressible and diagonalize the vector Laplacian, with eigenvalue $k^2+q^2$.

A complete basis of these modes with our defined BCs at $R$ is given by $k=\hat{k}_n/L$ and $q=\hat{q}_m/R$ where $\hat{k}_n=2n\pi$ for integer $n$ and with $\hat{q}_m$ the $m^{th}$ zero of $J_0(x)$.  We thus write the flow field in terms of these as 
\begin{equation}
\label{eq:sum}
\vec{\Delta}(\rho,z)=\frac{1}{LR}\sum\limits_{k}\sum\limits_{q} \left\{ \frac{k}{q}e^{i k z}J_1(q\rho), i e^{i k z}J_0(q\rho) \right\}  \tilde{\Delta}_{k,q}
\end{equation}
\section{Detailed Calculation of Susceptibilities}
\label{sec:calc}
To calculate the conjugate momentum 
we use that:
\begin{equation}
\begin{array}{r}
\int_0^L dz e^{i\tilde{k}_iz/L} e^{i\tilde{k}_jz/L} = L \delta_{\tilde{k}_i+\tilde{k}_j} \\
\int_0^R 2\pi \rho d\rho J_0(\tilde{q}_i\rho/R) J_0(\tilde{q}_j\rho/R) = \pi R^2 J_1(\tilde{q}_i)^2 \delta_{\tilde{q}_i-\tilde{q}_j} \\
= 2 R^2 / \tilde{q_i} \delta_{\tilde{q}_i-\tilde{q}_j}  =(2R/q) \delta_{\tilde{q}_i-\tilde{q}_j}

\end{array}
\end{equation}
Where in the last line we have used the asymptotic form for $J_1(x)$, assuming that $R$ is large.  From the above, the kinetic energy is given by:
\begin{equation}
\begin{array}{rl}
\mathcal{T} &=\frac{\rho_{3D}}{2} \int dz \int 2\pi\rho d\rho \left| \frac{\partial \vec{\Delta}(\rho,z)}{\partial t} \right|^2\\
&=\frac{\rho_{3D}}{2}\frac{1}{LR}\sum\limits_{k} \sum\limits_{q} -(1+k^2/q^2)\frac{2}{q}\dot{\tilde{\Delta}}_{k,q}\dot{\tilde{\Delta}}_{-k,q} 
\end{array}
\end{equation}
So that the conjugate momentum to $\tilde{\Delta}_{k,q}$ is given by 
\begin{equation}
\label{eq:pdef}
\tilde{p}_{k,q} =\frac{\partial \mathcal{T}}{\partial \dot{\tilde{\Delta}}_{k,q}}=-\frac{2\rho_{w}}{q RL} \dot{\tilde{\Delta}}_{-k,q},
\end{equation}
We immediately drop the factor of $(1+k^2/q^2)$ anticipating that relevant modes will have wavelengths much larger than the radius of the axon, so that $k<<q$.

We assume that the potential energy is stored at the membrane surface and as such can be written in terms of the fields: 
\begin{equation}
\label{eq:hcdef}
\begin{array}{rl}
\tilde{h}_k &=\frac{k}{R} \sum\limits_{q} \frac{1}{qr_0} J_1(qr_0) \tilde{\Delta}_{k,q}=\int dz e^{-ikz} h(z) \\
\tilde{l}_k &= \frac{-k}{R} \sum\limits_{q} J_0(qr_0)  \tilde{\Delta}_{k,q}=\int dz e^{-ikz} l(z) 

\end{array}
\end{equation}
where the relative height field, $h(z)$, and lateral stretch field, $l(z)$ are given by:

\begin{equation}
\begin{array}{rl}
h(z) &= \frac{1}{r_o} \Delta_\rho(r_0,z)\\
l(z) &=\frac{\partial}{\partial{z}} \Delta_z(r_0,z)
\end{array}
\end{equation}

Several possible potential energy terms that can be physically motivated can be written in terms of these surface modes.  The first one arises from interfacial energy and is written: 
\begin{equation}
\mathcal{U}_{interfacial} = 2\pi r_0 \sigma \int dz \frac{\partial h(z)}{\partial z}^2
\end{equation}
where surface tension $\sigma$ is an effective parameter that arises from a combination of membrane tension $\sigma_m$ and a cytoskeletal contribution $\sigma_c$, for example from fibers oriented in the z direction. However, in the regime where $kr_0 <<1$, elastic contributions dominate and so we do not consider this term in more detail.  Two of these elastic terms are easiest to motivate.  The first is a $2D$ bulk modulus which can be written:
\begin{equation}
\mathcal{U}_{stretch} = 2\pi r_0 \kappa_{2D} \int dz \frac{(h(z)+l(z))^2}{2}
\end{equation}
Where $\kappa_{2D}$ is the 2D bulk modulus of the membrane with possible contributions from the membrane proximal cytoskeleton.

Here we focus on a term which specifically punishes radial elastic deformations, motivated by experiments showing evenly spaced actin rings:  
\begin{equation}
\mathcal{U}_{radial}= 2\pi r_o \kappa \int dz \frac{h(z)^2}{2}
\end{equation}
where $\kappa$ arises primarily from cytoskeletal elements oriented perpendicularly to the axon.

For our potential energy we thus take:
  \begin{equation}
  \begin{array}{rl}
\mathcal{U} &= 2\pi r_o \kappa \int dz \frac{h(z)^2}{2} -h(z)F^h(z)-l(z)F^l(z)\\
&= 2\pi r_0 \kappa \sum_k \frac{\tilde{h}_k \tilde{h}_{-k}}{2} -\tilde{h}_k \tilde{F}^h_{-k}-\tilde{l}_k \tilde{F}^l_{-k} 
\end{array}
\end{equation}

where $F^{h/l}(z)$ and its transform $\tilde{F}^{h/l}_{-k} = \int dz e^{-ikz} F^{h/l}(z)$ are applied fields that couple to the fields $h$ and $l$.

The time derivatives of the momentum modes are given by:
\begin{equation}
\label{eq:dotp}
\dot{p}_{k,q}= -\frac{\eta}{\rho_{3D}}(k^2+q^2)p_{k,q} -\frac{\partial}{\partial \tilde{\Delta}_{k,q}} \mathcal{U}
\end{equation}
As $\mathcal{U}$ depends only on dynamic variables through the fields $l$ and $h$ we have that:
\begin{equation}
\label{eq:derivU}
\begin{array}{rl}
\frac{\partial}{\partial \tilde{\Delta}_{k,q}} \mathcal{U}=&\left(\frac{\partial \tilde{h}_k}{\partial \tilde{\Delta}_{k,q}} \right) \frac{\partial}{\partial \tilde{h}_{k}} \mathcal{U}+\left(\frac{\partial \tilde{l}_k}{\partial \tilde{\Delta}_{k,q}} \right) \frac{\partial}{\partial \tilde{l}_{k}} \mathcal{U}\\
=&\frac{k}{RL}\frac{1}{qr_0} J_1(qr_0)\left(2\pi r_0 \kappa \tilde{h}_{-k} +\tilde{F}^h_{-k}\right)  \\
&-\frac{k}{RL}J_0(qr_0) \tilde{F}^l_{-k}   
\end{array}
\end{equation}
Using equation~\ref{eq:pdef} we can remove the $p$s from equation~\ref{eq:dotp} to give:
\begin{equation}
\label{eq:notF}
\begin{array}{l}
\frac{2\rho_{3D}}{qR}\left(\ddot{\tilde{\Delta}}_{-k,q} +(\eta/\rho_{3D})q^2\dot{\tilde{\Delta}}_{-k,q}\right) \\
=\frac{k}{R}\frac{1}{qr_0} J_1(qr_0)\left(2\pi r_0 \kappa \tilde{h}_{-k} +\tilde{F}^h_{-k}\right)-
\frac{k}{R}J_0(qr_0)\tilde{F}^l_{-k}
\end{array}
\end{equation}

It is now convenient to switch to Fourier space in time, defining for all dynamic variables $A$
\begin{equation}
\begin{array}{rl}
A_\omega&= \int dt e^{i\omega t} A(t)\\
A(t)&=\frac{1}{2\pi}\int d\omega e^{-i\omega t} A_\omega
\end{array}
\end{equation}

In Fourier space, Eq ~\ref{eq:notF} becomes:

\begin{equation}
\label{eq:Deltakqw}
\tilde{\Delta}_{k,q,\omega}=\frac{kq}{2\rho_{3D}}\frac{\frac{1}{qr_0} J_1(qr_0)\left(2\pi r_0 \kappa \tilde{h}_{k,\omega} +\tilde{F}^h_{k,\omega}\right)-J_0(qr_0)\tilde{F}^l_{k,\omega}}{\omega^2+i\omega (\eta/\rho_{3D})q^2}.
\end{equation}

We can sum both sides of the equation and replace the $\Delta$s with $h$s using eq.~\ref{eq:hcdef}:

\begin{equation}
\begin{array}{l}
\tilde{h}_{k,\omega}=\frac{k^2}{2\rho_{3D} R} \times \\
\sum\limits_{q}q \frac{\left(\frac{1}{qr_0} J_1(qr_0)\right)^2\left(2\pi r_0 \kappa \tilde{h}_{k,\omega} +\tilde{F}^h_{k,\omega}\right)-\frac{1}{qr_0} J_1(qr_0)J_0(qr_0)\tilde{F}^l_{k,\omega}}{\omega^2+i\omega (\eta/\rho_{3D})q^2} 
\end{array}
\end{equation}

We can now take $R \rightarrow \infty$.  The density of Bessel Function zeros is given by $\Delta q = \pi/R$ so that $1/R\sum\limits_{q} \rightarrow \int \frac{dq}{\pi}$ yielding

\begin{equation}
\begin{array}{l}
\tilde{h}_{k,\omega}=\frac{k^2}{2\pi \rho_{3D} } \times \\
\int qdq \frac{\left(\frac{1}{qr_0} J_1(qr_0)\right)^2 \left(2\pi r_0 \kappa \tilde{h}_{k,\omega} +\tilde{F}^h_{k,\omega}\right)-\frac{1}{qr_0} J_1(qr_0)J_0(qr_0)\tilde{F}^l_{k,\omega}}{\omega^2+i\omega (\eta/\rho_{3D})q^2} 
\end{array}
\end{equation}

Transforming to coordinate $x=qr_o$ gives:

\begin{equation}
\tilde{h}_{k,\omega} = \frac{k^2M_{11}(\alpha)}{2\pi \rho_{3D} r_0^2 \omega^2}\left(2\pi r_0 \kappa \tilde{h}_{k,\omega} +\tilde{F}^h_{k,\omega}+\frac{M_{12}(\alpha)}{M_{11}(\alpha)}\tilde{F}^l_{k,\omega}\right),
\end{equation}
where 
\begin{equation}
\alpha=\frac{\rho_{3D} r_0^2 \omega}{\eta}
\end{equation}
is a Reynolds-like number of the mode and where the $M(\alpha)$s are dimensionless functions given by:
\begin{equation}
\begin{array}{rl}
M_{11}(\alpha) &=\int_{0}^{\infty}  \frac{dx}{x}\frac{J_1(x)J_1(x)}{1+i x^2/\alpha} \\
M_{12}(\alpha) &=-\int_{0}^{\infty}  dx \frac{J_1(x)J_0(x)}{1+i x^2/\alpha} \\
M_{22}(\alpha) &=\int_{0}^{\infty}  x dx \frac{J_0(x)J_0(x)}{1+i x^2/\alpha} \\
\end{array}
\end{equation}
This allows us to write:
\begin{equation}
\tilde{h}_{k,\omega}=\chi^{hh}_{k,\omega}F^h_{k,\omega}+\chi^{hl}_{k,\omega}F^l_{k,\omega}
\end{equation}
with
\begin{equation}
\begin{array}{rl}
(\chi^{hh}_{k,\omega})^{-1} &= \frac{2\pi \rho_{3D} r_0^2 \omega^2}{k^2M_{11}(\alpha)}-2\pi r_0 \kappa \\
(\chi^{hf}_{k,\omega})^{-1} &= \frac{2\pi \rho_{3D} r_0^2 \omega^2}{k^2M_{12}(\alpha)}-\frac{M_{11}(\alpha)}{M_{12}(\alpha)}2\pi r_0 \kappa
\end{array}
\end{equation}

\end{article}

\end{document}